\documentclass[twocolumn,a4paper,showpacs,aps,pra] {revtex4-1}

\usepackage{graphicx,color}

\begin{document}

\title{Photoemission spectroscopy with high-intensity short-wavelength lasers}
\author {Song Bin Zhang and Nina Rohringer}
\affiliation {Max Planck Institute for the Physics of Complex Systems, 01187 Dresden, Germany\\
              Center for Free-Electron Laser Science (CFEL), 22761 Hamburg, Germany}
\email {\color{blue}{song-bin.zhang@pks.mpg.de \\
        nina.rohringer@pks.mpg.de}}
\date{September 15, 2013}
\revised{\today}

\begin{abstract}
We theoretically study the process of photoelectron emission of the Helium atom by a high-intensity short-wavelength laser at a
resonance condition of the residual singly charged ion. Photoionization followed by strong resonance coupling in the ion leads to
a change in the photoelectron spectrum due to Rabi oscillations in the residual ion. Similarly to resonance fluorescence at high
laser intensity, the photoelectron spectrum at high intensities evolves into a multi-peaked structure. The number of peaks in the 
photoelectron spectrum is related to the number of Rabi cycles following the photoionization process. Moreover, the strong laser-induced coupling to nonresonant states of the residual ion have an imprint on the photoelectron spectrum, leading to additional, isolated peaks at the lower or higher energy sides. The effect should be observable at current seeded XUV free-electron lasers and persists after volume integration in a realistic experimental geometry.
\end{abstract}

\maketitle

The photoelectric effect is typically the most likely process induced by a short-wavelength radiation impinging on an atomic, molecular or solid target. At short-wavelength synchrotron sources, angular and energy-resolved photoelectron spectroscopy {\color{blue} \cite{Turner1970,Stefan2003,Stefan2007}} is an indispensable tool to map the electronic bandstructure of solids, surfaces and interfaces. Intensities at synchrotrons are low, so that photoionization can be treated by perturbation theory, \textit{i.e.}, can be considered as a single-photon process.
At long optical wavelengths multiphoton ionization is a well-studied topic and lies at the core of present day optical strong field physics {\color{blue} \cite{Blanchet1999,Paulus2001,Stolow2003,Stolow2004,Corkum2007,Blaga2009,Krausz2009,Ott2013}}. With the invention of free-electron lasers, resulting in a tremendous boost in intensities, nonlinear coupling of short-wavelength radiation with matter becomes possible {\color{blue} \cite{Rohringer2008,Kanter2011,Rohringer2012,Doumy2011,Demekhin2011b,Demekhin2013a}} and sequential multiple ionization can be a dominant process {\color{blue} \cite{Rohringer2007,young2010,Rudek2012}}.\\

There are several ways, a high intensity short-wavelength field can alter the photoionization process. A nonresonant radiation field of high intensity can dress the single-electron continuum states in atoms {\color{blue} \cite{Demekhin2012b}} or ions {\color{blue} \cite{Toyota2007,Toyota2008}}, predicted to result in a multi-peak structures of the photoelectron spectra. Likewise, dressing of the continuum can lead to multi-peaked electron spectra of autoionizing states {\color{blue} \cite{Rzazewski1983}}. On the other hand, strong resonant coupling of bound-to-bound transitions leads to the well-known Mollow triplets in resonance fluorescence {\color{blue} \cite{Mollow1969,Robinson1986}} and similarly multi-peaked and broadened resonant Auger spectra {\color{blue} \cite{Rohringer2008,Kanter2011,Rohringer2012,Demekhin2011b,Demekhin2013a}}, resonant multiphoton ionization {\color{blue} \cite{Rogus1986,Meier1994}} and sequential two-photon ionization {\color{blue} \cite{Demekhin2012a}}. The multi-peaked spectra can typically be explained in terms of the dressed state picture, or by dynamical interference in the emission process {\color{blue} \cite{Demekhin2012a}}.\\

In this work, we study yet another field-dressing effect, which will have an imprint on photoelectron spectra in a high-intensity short-wavelength field and is timely in view of the first operating seeded short-wavelength free-electron laser (FEL) {\color{blue} \cite{Allaria2012a,Allaria2012b}}. We suppose that a focused FEL beam is impinging on a gas target of initially neutral Helium atoms. The FEL is resonantly tuned to the 1s-$n$p resonance of the Helium ion, so that subsequent to single-photon ionization event, the residual ion undergoes strong resonant interaction with the FEL radiation (see Fig. {\color{blue} 1} for the level system of He and He$^+$), \textit{i.e.}, inducing Rabi oscillations in the ion. The periodic modulation of the residual ionic states will have an imprint on the photoelectron spectra. With rising FEL intensity the photoelectron peak is predicted to evolve into a multi-peaked structure. Moreover, nonresonant states of the ion will lead to additional peaks at lower or higher energies. To demonstrate the effect, we study the Helium atom. The results, however, are generally valid and will apply to other atomic and molecular systems. Unless otherwise stated, atomic units (a.u.) are used throughout the paper.\\  
  
We consider He initially in its ground electronic state $|1\textit{s}^{2}\rangle$ of energy $E_{1s^{2}}$, which is ionized into the ground state $|1\textit{s}\rangle$ of He$^{+}$ by absorbing one photon of frequency $\omega$. Thereby an intermediate single-electron continuum state $|1\textit{s},\varepsilon\rangle$ with energy $E_{1s}+\varepsilon$ is occupied, where $E_{1s}$ is the energy of residual He$^{+}(1s)$ ion and $\varepsilon$ is the kinetic energy of the photo-electron; subsequently $|1\textit{s},\varepsilon\rangle$ is resonantly coupled to the $|n\textit{p},\varepsilon\rangle$ state (see Fig. {\color{blue} 1}), supposing the photon energy exactly resonant with the energy between states $|1\textit{s}\rangle$ and $|n\textit{p}\rangle$. $E_{np}$ is the energy of He$^{+}(np)$, the energy of the two-particle state $|n\textit{p},\varepsilon\rangle$ is hence $E_{np}+\varepsilon$. We suppose a linearly polarized electric field $G=g_{0}g(t)cos(\omega t)$ with electric field strength $g_{0}$ and pulse envelope $g(t)$. The total time dependent wavefunction can be expanded in

\begin{eqnarray}
\Psi(t)=a_{1s^{2}}(t)|1\textit{s}^{2}\rangle+
\int a_{1s, \varepsilon}(t)e^{-i\omega t}|1\textit{s},\varepsilon\rangle d\varepsilon   \nonumber \\ 
+\sum_n \int a_{np, \varepsilon}(t)e^{-2i\omega t}|n\textit{p},\varepsilon\rangle d\varepsilon,
\end{eqnarray}
where $a_{1s^{2}}(t)$, $a_{1s, \varepsilon}(t)$ and $a_{np, \varepsilon}(t)$ are the time-dependent amplitudes of the levels $|1\textit{s}^{2}\rangle$,~$|1\textit{s},\varepsilon\rangle$ and $|n\textit{p},\varepsilon\rangle$, respectively. Since the field is supposed to be strong, nonresonant coupling terms become important and residual ionic states beyond the resonant excitation have to be taken into consideration. In the following calculations, we have included six final states  $n=2,...,7$ (Fig. {\color{blue} 1}), which resulted in converged results. The field is supposed to be either resonant with the 1s-2p or 1s-3p transition in the residual ion. In the ansatz of the wavefunction we separated a rapid-evolving phase factor $e^{-i\omega t}$ or $e^{-2i\omega t}$ {\color{blue} \cite{Demekhin2011b,Demekhin2012a,Demekhin2012b}}. \\

\begin{figure}[h]
\includegraphics[width=7.5 cm]{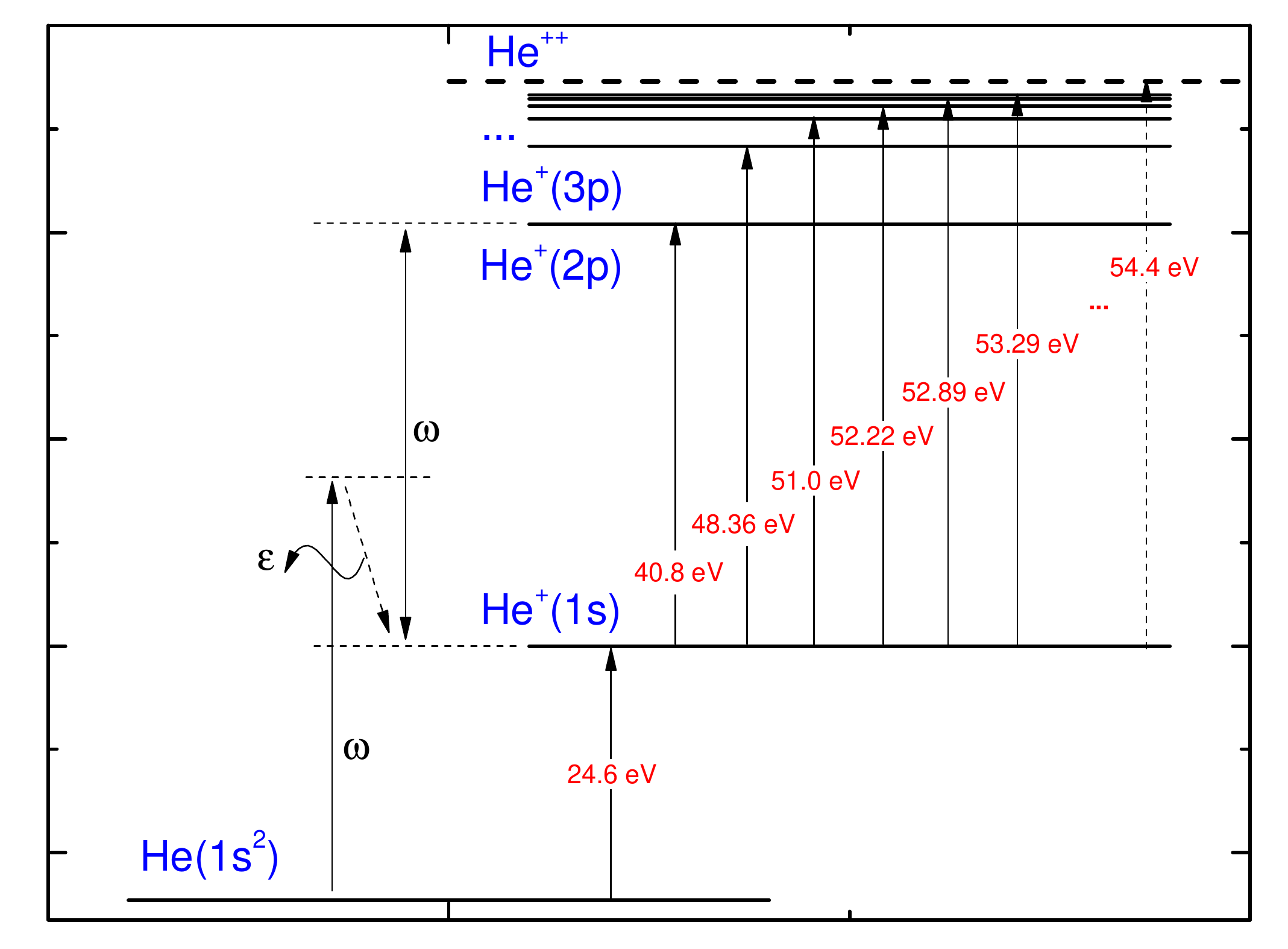}
\caption{Energy levels of He and He$^{+}$ and the schematics of photoionization process with resonant coupling in the residual ion}
\end{figure}

Inserting the total wavefunction into the time-dependent Schr\"odinger equation for the total Hamiltonian and implying the rotating wave approximation {\color{blue} \cite{Gamaly2011,Shore2011}} and the local approximation {\color{blue} \cite{Cederbaum1981,Domcke1991,Pahl1996,Demekhin2011b}} leads to a set of equations determining the evolution of the expansion coefficients:

\begin{eqnarray}\
i\dot a_{1s^{2}}(t)={\LARGE [}E_{1s^{2}}-\frac{i\Gamma(\varepsilon)}{2}g^{2}(t){\LARGE ]}a_{1s^{2}}(t) 
~ ~ ~ ~ ~ ~ ~ ~ ~ ~ ~ ~ ~ ~ ~ ~ ~ ~ ~ ~ ~ ~ ~ ~ ~ ~ ~ ~\nonumber \\
+\frac{D_{1s^{2}}^{\dagger}(\varepsilon)g_{0}}{2}g(t)a_{1s,\varepsilon}(t),  \nonumber \\
i\dot a_{1s,\varepsilon}(t)=\frac{D_{1s^{2}}(\varepsilon)g_{0}}{2}g(t)a_{1s^{2}}(t)+(E_{1s}+\varepsilon
-\omega)a_{1s,\varepsilon}  ~ ~ ~ ~ ~ ~ ~\nonumber \\
+\frac{d_{np}^{\dagger}g_{0}}{2}g(t)a_{np,\varepsilon}(t),\nonumber \\
i\dot a_{np,\varepsilon}(t)=\frac{d_{np}g_{0}}{2}g(t)a_{1s,\varepsilon}(t)
+(E_{np}+\varepsilon-2\omega)a_{np,\varepsilon}. ~ ~ ~ ~ ~ ~ ~ ~ ~
\end{eqnarray}
Here $D_{1s^{2}}(\varepsilon)=\langle 1\textit{s},\varepsilon|z|1\textit{s}^2 \rangle$ are the energy dependent dipole transition matrix elements to the continuum states $|1\textit{s},\varepsilon\rangle$. Absorption of a photon of energy $\omega=40.800$ eV (1s-2p transition of He$^{+}$) leads to a continuum state with $D_{1s^{2}}=0.502$ a.u., whereas for $\omega=48.356$ eV (1s-3p transition of He$^{+}$) $D_{1s^{2}}=0.387$ a.u. {\color{blue} \cite{Yan1998}}. We, however, suppose that the transition dipole is constant around the peak-electron energy, which is a good approximation. 
$ d_{np}= \langle n\textit{p},\varepsilon|z|1\textit{s}, \varepsilon \rangle = \langle n\textit{p}|z|1\textit{s} \rangle$ are the dipole transition matrix elements between the residual ionic states $|1\textit{s}\rangle$ and $|n\textit{p}\rangle$. They are independent of the photon energy, and are $ d_{np}= $ $0.37323$, $0.14949$, $0.08813$, $0.06039$, $0.04474$, and $0.03513$ a.u. for n=$2$,~$3$,~$4$,~$5$,~$6$, and $7$, respectively. $\Gamma(\varepsilon)=2\pi|D_{1s^{2}}(\varepsilon)g_{0}/2|^{2}$, and $\frac{i\Gamma(\varepsilon)}{2} g^{2}(t)$ represent the ''dynamic'' ionization rate from the He ground state.\\

The electron spectrum pertaining to the final ionic channel  $|nl\rangle$ is given by

\begin{equation}
\sigma_{nl}(\varepsilon)=\lim _{t\rightarrow \infty} {\langle a_{nl,\varepsilon}(t)|a_{nl,\varepsilon}(t) \rangle}.
\end{equation}
The probabilities of occupying a final ionic channel $|nl\rangle$ are determined by
\begin{equation}
p_{nl}=\int {\sigma_{nl}(\varepsilon)} d\varepsilon.
\end{equation}
The total electron spectrum can be computed as the incoherent sum of the electron spectra to every final state {\color{blue} \cite{Demekhin2011a}}
\begin{equation}
\sigma(\varepsilon)=\sum _{nl}\sigma_{nl}(\varepsilon).
\end{equation}

The system of Eq.({\color{blue} 2}) was solved numerically employing a Gaussian pulse $g(t)=e^{-4ln(2)t^{2}/\tau^{2}}$ of $\tau=$ 30 fs duration (full width half maximum of the electric field). Note that the spectral bandwidth of the pulse is $\sim$ 0.02 eV and small compared to the typical features in the electron spectra.

\begin{figure}[h]
\includegraphics[width=8 cm]{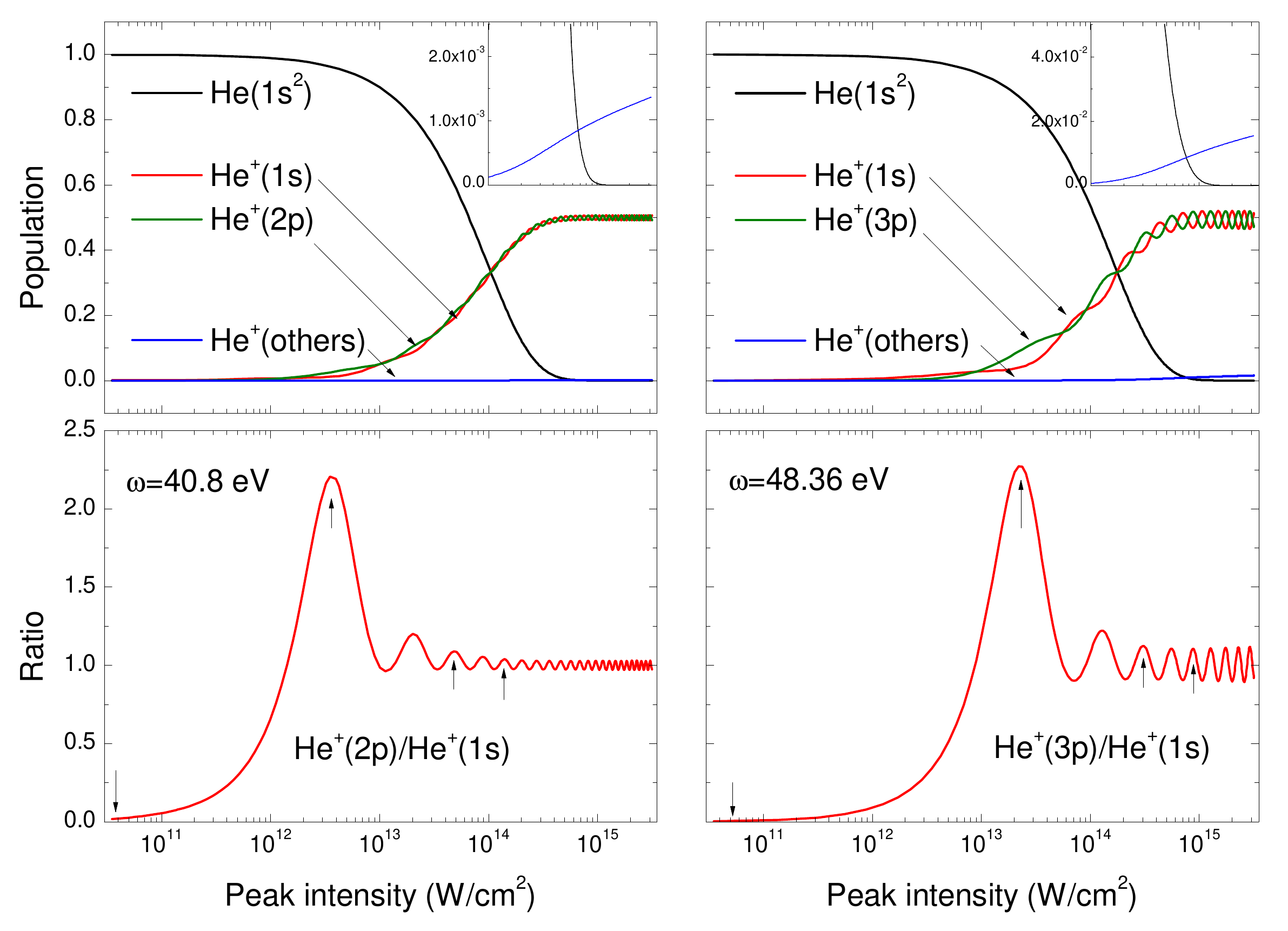}
\caption{Top panels show the final populations of the ground state He $(1s^{2})$, the ionic ground state He $^{+}(1s)$ and ionic excited states  He$^{+}(2p)$ and He$^{+}(3p)$ as a function of the applied peak intensity.
The lower panels shows the ratio of the final populations of the two resonantly coupled states as function of the applied peak intensity. The vertical lines indicate the peak intensities for which the photoemission spectra are depicted in Fig.\ {\color{blue} 3} and Fig.\ {\color{blue} 5}.}
\end{figure}

Fig. {\color{blue} 2} shows the populations of the atomic ground state $|a_{1s^{2}}|^{2}$ and ionic states $p_{nl}$ after interaction with the laser pulse as a function of the peak intensity (1 a.u. intensity = $3.51\times10^{16}$ W/cm$^{2}$) for two different frequencies $\omega=$ 40.8 and 48.356 eV. 
For intensities beyond $\sim 3\times 10^{12}$ W/cm$^{2}$, the populations of the two resonantly coupled ionic states exhibit pronounced Rabi oscillations. This can be seen in the relative occupation of the final occupations of the two resonantly coupled states, which shows oscillations, as shown in the lower panel of Fig. {\color{blue} 2}. The minima in the occupation ratio correspond to intensities of $n\pi$ pulses, i.e.\ completion of $n$ complete Rabi cycles during the interaction with the laser pulse. Note that the transition dipole in the case of $\omega$=40.8 eV is larger than that of case $\omega$=48.356 eV, which results in more Rabi oscillations as shown in Fig. {\color{blue} 2}. The total populations for other ionic states (n $\geq$ 3 and n=2,4--7) are quite small, the total populations of those states is about 0.07$\%$ and 0.6$\%$ for a peak intensity of $5\times10^{14}$ W/cm$^{2}$ for the cases $\omega$=40.8 eV and $\omega$=48.356 eV, respectively; those numbers are about 0.14$\%$ and 1.5$\%$, respectively, when the peak intensity reaches about 0.1 a.u. ($3.51\times10^{15}$ W/cm$^{2}$). Despite their small final-state occupation, these states play an important role on the shape of the photoelectron spectrum.  

\begin{figure}[h]
\includegraphics[width=8 cm]{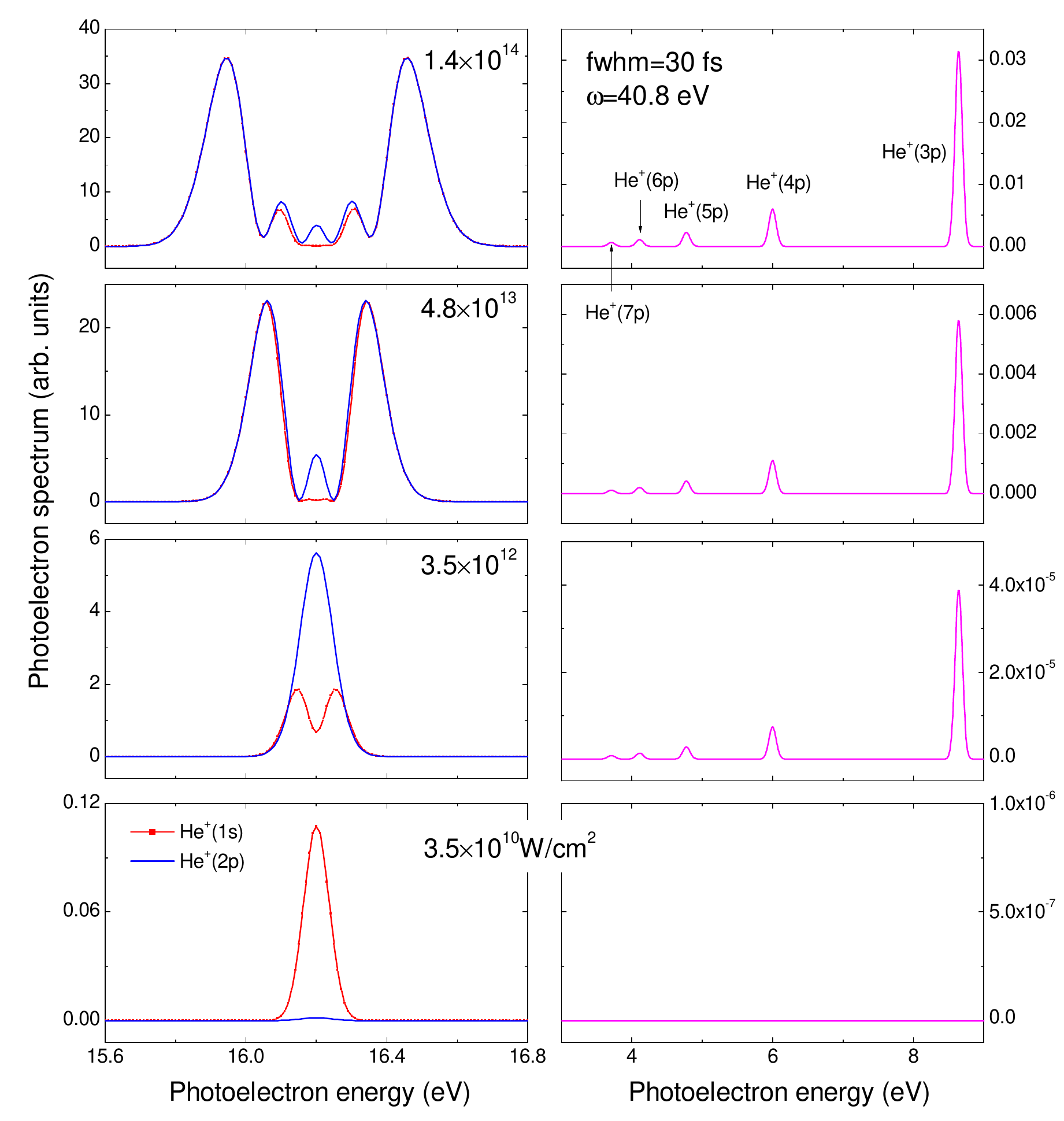}
\caption{Shown are the photo-electron spectra for different peak intensities marked by arrows in the lower panels of Fig. {\color{blue} 2} for the case of $\omega$=40.8 eV (resonance condition for the residual ionic states $|1\textit{s}\rangle$ and $|2\textit{p}\rangle$). The left panels show the main emission line corresponding to the ionic final state $|1\textit{s}\rangle$, the right panels correspond to photoelectron emission with excited ionic final states.}
\end{figure}

The photoelectron spectra pertaining to different ionic states for different field peak intensities (marked by arrows in Fig. {\color{blue} 2}) for $\omega$=40.8 eV are shown in Fig. {\color{blue} 3}. The spectrum of the lowest intensity case ($3.5\times10^{10}$ W/cm$^{2}$) is a Gaussian curve of spectral width in correspondence to the bandwidth of the applied pulse. The residual ion is in the state $|1\textit{s}\rangle$ and the spectrum is centered around $\varepsilon=\omega+E_{1s^{2}}-E_{1s}=16.2$ eV, as expected in the low-intensity case. As the field intensity increases, the ionic system undergoes Rabi oscillations between the two resonant ionic states $|1\textit{s}\rangle$ and $|2\textit{p}\rangle$. This results in the build-up of multi-peak structures in the spectrum, resulting from the dynamical Stark effect of the residual ionic states {\color{blue} \cite{Toyota2007,Toyota2008,Demekhin2012b,Demekhin2012a}}; the main peak shows an inversion symmetry at around 16.2 eV. With increasing field intensity and completion of a higher number of Rabi cycles, the number of maxima and their energy splitting (a measure of the averaged Rabi frequency) increases. At a field intensity of $3.5\times10^{12}$ W/cm$^{2}$ the residual ion undergoes half a Rabi cycle ($\pi /2$ pulse) and the spectra pertaining to the final ionic states $|2\textit{p}\rangle$ and $|1\textit{s}\rangle$ exhibit a single and two peaks, respectively. The ratio between their populations is about 2.2, the maximum value for all the field intensities; For field intensities $4.8\times10^{13}$ and $1.4\times10^{14}$ W/cm$^{2}$, the ionic system completes 2.5 and 4.5 Rabi cycles, respectively, and the spectra exhibit more complex structures.\\ 

With the field intensities increasing, the non-resonant final states start to contribute to the total spectrum and manifest themselves by small peaks at the low energy side of the main photoelectron peak (shown in the right panels of Fig. {\color{blue} 3}). Explicitly, the peaks around 3.7, 4.1, 4.8, 6.0 and 8.7 eV appear, corresponding to contributions from states $|7\textit{p}\rangle$, $|6\textit{p}\rangle$, $|5\textit{p}\rangle$, $|4\textit{p}\rangle$ and $|3\textit{p}\rangle$, respectively; these non-resonant states become more and more important and their spectral intensity increases with increasing field intensity. Note that with increasing field intensity, the small peaks shift to lower energy region. Another consequence of the nonresonant coupling is a small asymmetry in the main photoelectron peaks, \textit{i.e.}, the higher energy region is more populated than the lower energy region.\\

\begin{figure}[h]
\includegraphics[width=8 cm]{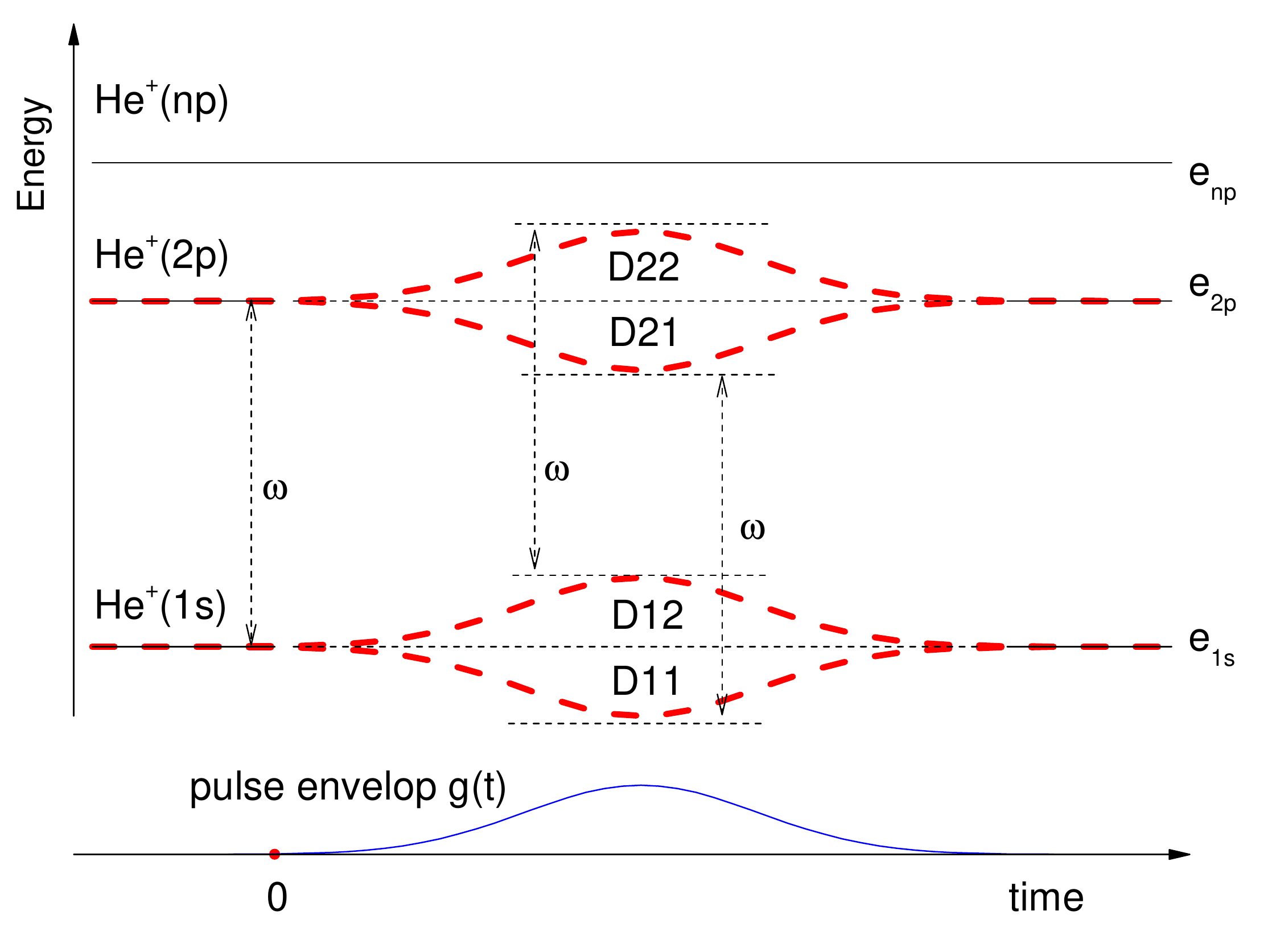}
\caption{Schematics of the energy levels in the strong laser field. States $|1\textit{s}\rangle$ and $|2\textit{p}\rangle$ are resonantly coupled by the intense laser of frequency $\omega$ and become dressed (states D11, D12, D21 and D22). State $|n\textit{p}\rangle$ is not dressed by the field.}  
\end{figure}

These features can be easily explained within the dressed state model between the ionic states $|1\textit{s}\rangle$ and $|2\textit{p}\rangle$ and the off-resonant transition between states $|1\textit{s}\rangle$ and $|n\textit{p}\rangle$ (see Fig. {\color{blue} 4}). When the field is on, states $|1\textit{s}\rangle$ and $|2\textit{p}\rangle$ become dressed, split into states  D11, D12 and D21, D22, respectively. 
Their energy splitting D12-D11 (D22-D21) is given by the Rabi frequency, which adiabatically follows the intensity envelope. 
The transition from $|1s\rangle$ to $|np\rangle$ is far from the resonance condition, and at the considered field intensities, 
dressing of the $|np\rangle$ state  can be neglected, \textit{i.e.}, the energy level ofthe  $|np\rangle$ state can be considered as constant. At initial times of the laser pulse, the detuning of the  $|np\rangle$ state from the undressed state  $|1s\rangle$ is large. As the laser intensity ramps on, the  $|1s\rangle$ state evolves into the dressed states D11 and D12, resulting in smaller (larger) detuning of the $|np\rangle$ state  from the D12 (D11) state. This leads to a higher (lower) transition probability to the $|np\rangle$ state, which results in a small asymmetry in the main photoelectron feature (at roughly 16.2 eV), precisely, slightly higher photoelectron yield at the higher energy side (The asymmetry is too small to be visible in Fig. {\color{blue} 3}, comparing the heights of the main peaks a change of 0.2 is observed, which translates into a relative change of only 0.5$ \% $). Increasing the peak intensity, therefore results in an increase of the photoelectron yield assigned to the residual ionic state $|np\rangle$, along with a small energy shift towards the lower energy side (The shift is hardly visible in Fig. {\color{blue} 3} and is of the order of 0.001 eV for the intensity of $3.5\times10^{12}$ compared to $1.4\times10^{14}$ W/cm$^2$).\\

The situation slightly changes, if we consider strong resonant coupling of $|1\textit{s}\rangle$ and $|3\textit{p}\rangle$ levels ($\omega$=48.356 eV). Photoelectron spectra for that case are shown in Fig. {\color{blue} 5}. The main spectrum is centered around 23.756 eV and shows similar changes with increasing peak intensity as for the case of case $\omega$=40.8 eV. The peaks in the spectra around 18.8, 19.2, 19.9 and 21.1 eV are contributions from states $|7\textit{p}\rangle$, $|6\textit{p}\rangle$, $|5\textit{p}\rangle$ and $|4\textit{p}\rangle$, respectively. In that case, however, the $|2\textit{p}\rangle$ level lies in between the field-dressed states, resulting in an additional peak at the higher energy side of the main photoelectron feature (at about 31.3 eV). As the field intensity increases, the main spectra show more significant asymmetric patterns, which reflects the role of the non-resonant states. Omitting the contribution of nonresonant states would result in symmetric emission patterns. The peaks pertaining to states $|7\textit{p}\rangle$, $|6\textit{p}\rangle$, $|5\textit{p}\rangle$ and $|4\textit{p}\rangle$ gradually shift to lower energy, while the peak pertaining to final state state $|2\textit{p}\rangle$ gradually shifts to higher energy, a feature which is consistent with the dressed-state picture introduced in Fig. {\color{blue} 4}.\\ 

\begin{figure}[h]
\includegraphics[width=8 cm]{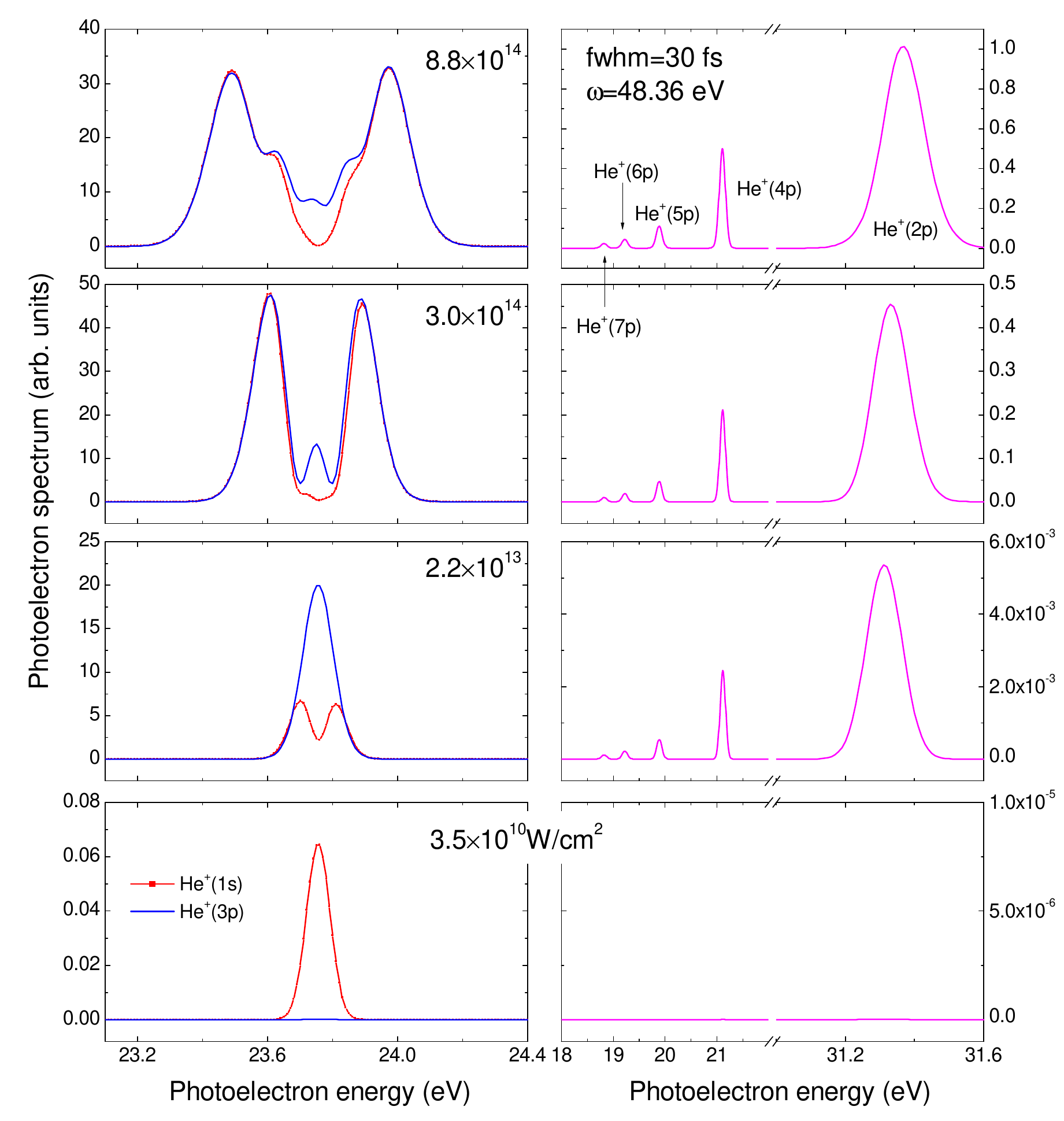}
\caption{ Same as in Fig. {\color{blue} 3}, but for the case of $\omega$=48.356 eV (resonance condition for ionic states $|1\textit{s}\rangle$ and $|3\textit{p}\rangle$).}
\end{figure}

\begin{figure}[t]
\includegraphics[width=8 cm]{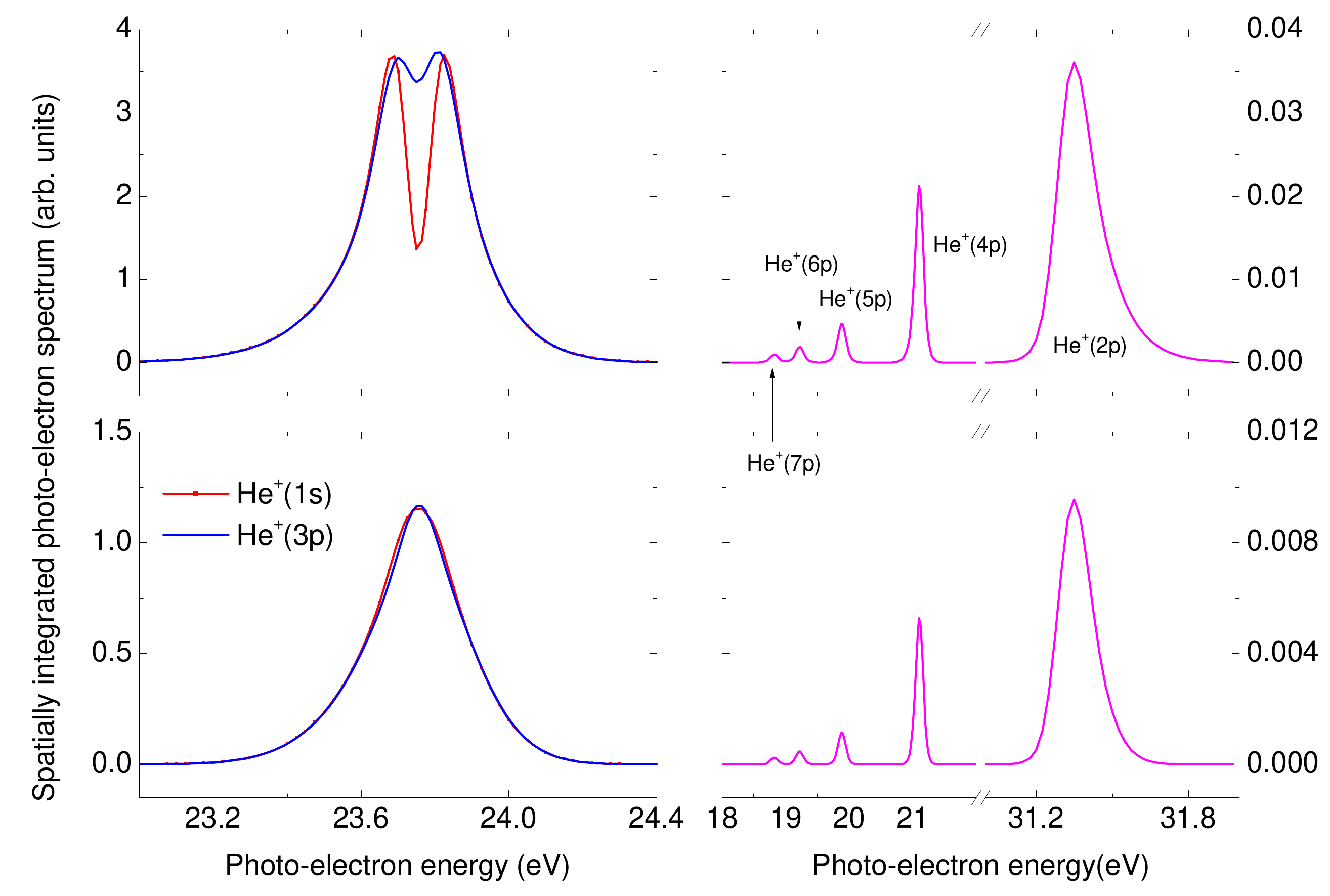}
\caption{Integrated photoelectron spectra for a spatially Gaussian beam profile, supposing a photon energy of $\omega$=48.356 eV, a focal radius of 0.53 $\mu$m, a cylindrical interaction region of 2.0 mm diameter, and pulse energies of about 7.0 $\mu$J (upper panel) and 0.7 $\mu$J (lower pannel).}
\end{figure}
So far, the discussed results applied to a single atom, experiencing a well-defined temporal field intensity. In a real experimental condition, the spatial intensity distribution of the laser pulse has to be taken into consideration. The spectra resulting from atoms sitting at different positions in the laser focus have to be integrated with the according weight. Spatially integrated photo-electron spectra for $\omega$=48.356 eV are shown in Fig. {\color{blue} 6}, assuming a spatial Gaussian beam profile with a focal radius of 0.53 $\mu$m and cylindrical gas target of 2.0 mm diameter. We compare results for a pulse energy of about 7.0 and 0.7 $\mu$J, corresponding to maximum peak intensities $3.51\times 10^{16}$ and $3.51\times 10^{15}$ W/cm$^{2}$, respectively. Comparing the low and high-intensity case, a pronounced difference in the photoelectron spectrum is observed, \textit{i.e.}, the described effects could be experimentally seen, despite the spatial average, which typically smears out features resulting from the high-intensity areas of the pulse. Splitting of the photoelectron line and the asymmetry, as well as the low-energy structures should hence be observable in real experimental conditions.
Note that the ionization probability averaged over the focal volume is only $\sim$ 3$\%$ and 10$\%$ for pulse energies of $\sim$ 0.7 and 7.0 $\mu$J, respectively. The parameters of this study were chosen within the accessible range of the first seeded FEL, (FERMI at Elettra  {\color{blue} \cite{Allaria2012a,Allaria2012b}}. Typical pulse duration achievable at this source range from $\sim$ 30-100 fs, with pulse energies of $\sim$ 20-30 $\mu$J and photon energies of $\sim$ 20-100 eV. Our predicted effect should therefore be readily observable at present day short-wavelength FEL sources.  \\

In summary, we have presented a theoretical study of short-wavelength photoelectron spectroscopy of He at high laser intensity and resonant conditions in the residual ion. Dynamical, field-induced dressing of the residual ion results in strong modification of the photoelectron emission spectrum, featuring Autler-Townes {\color{blue} \cite{Autler1955}} like splitting of the emission line. Moreover, mixing of non-resonantly coupled states results in additional final ionic channels of excited states resulting in additional photo-emission peaks at lower or higher photon energies. Moreover, coupling to theses non-resonant states by the strong field results in slight asymmetries of the main emission line shape. With the fast development of seeded short-wavelength FEls, such nonconventional field-dressed photoemission spectra should be observable. The predicted effect was discussed in the Helium atom, but should be generally present in other atoms and molecules. Molecular systems will, however, feature more complex spectra, with imprints of nucelar wave-packet dynamics and will be studied in the near future. We hope that this work inspires future experiments at FEL sources.


%

\bibliographystyle{apsrev4-1}

\end{document}